# Electronic and magnetic structures and bonding properties of Ce$_2$CrN$_3$ and U$_2$CrN$_3$ from first principles

Samir F. Matar[a,*], Charbel N. Kfoury[b]

[a]CNRS, ICMCB, Université de Bordeaux. Pessac. France

[b]Faculté de Génie, Université Libanaise, Roumieh, Liban.

*Corresponding author's e_mail. Samir.Matar@icmcb.cnrs.fr ; abouliess@gmail.com

**Abstract.**

The electronic and magnetic structures of A$_2$CrN$_3$ (A = Ce, U) ternary compounds calculated based on band magnetism within DFT exhibit different behaviors of the nf elements (n = 4, 5 resp.). Charge analysis allows to formally express the two compounds as [A$_2$Cr]$^{\sim 5+}$ [N$_3$ ]$^{\sim 5-}$ thus classifying them as covalent nitrides, i.e. far from formal exchange of ±9 electrons. From establishing the energy-volume equations of state, the two compounds are found with hardness magnitudes: B$_0$(A=Ce) =192 GPa and B$_0$(A=U) = 243 GPa, within range of oxides due to covalent metal-nitrogen bonding shown as based on overlap matrix analysis. The uranium compound is harder due to a smaller volume and less compressible U versus Ce metals. Ce$_2$CrN$_3$ exhibits large magnetization on Cr (1.94μ$_B$) and a very small moment develops on cerium (0.14μ$_B$) pointing out to an intermediate valence state while in U$_2$CrN$_3$, M(Cr) = 0.49 μ$_B$ and M(U) = 0.97 μ$_B$. These results are stressed by broad band-like density of states (DOS) behavior for A=U and localized DOS for A=Ce. Both compounds are found ferromagnetic in the ground state.

## 1. Introduction

In last decades there has been a large interest in nitride compounds within the communities of Materials Science consisting of chemists, physicists and researchers in computational materials science. Such interest arises from the belonging of nitrides to different classes of compounds with a broad range of properties ranging from antiperovskite $TFe_3N$ (T = Fe, Mn, Ni, Pd, Pt ) exhibiting magnetic properties close to those of Invar alloy ($Ni_{0.35}Fe_{0.65}$) [1]; to binary, ternary and quaternary nitride semi-conductors ex. B-Al-Ga-N (cf. [2] and therein cited works) and to ultra-hard nitrides likely to replace diamond in applications, ex. $C_3N_4$, $C_{11}N_4$ [3], $BC_2N$ [4]; ….

Regarding *n*f based nitrides: $n$ = 4 for *Ln* (lanthanides) and 5 for *Ac* (actinides) a few ternary compounds were identified and studied as $Ce_2TN_3$ (T = Cr, Mn) [5, 6] and $A_2TN_3$ (A = Th, U) [7]. Focusing exemplarily on a lanthanide and an actinide, the isostructural ternaries $A_2CrN_3$ (A = Ce, U) crystallize in the orthorhombic *Immm* space group (SG). The structure sketched in Fig. 1 exhibits resemblance with those of tetragonal (*I4/mmm* SG) $K_2NiF_4$ archetype and $U_2IrC_2$. It can be noticed the change in the stoichiometry within the anionic sublattice: its filling up to 3 makes it intermediate between 4 in $K_2NiF_4$ and 2 in $U_2IrC_2$. Then the symmetry lowering from tetragonal to orthorhombic is due to this partial filling of the anionic substructure.

From the crystal chemistry standpoint, the A atoms are in the environment of nitrogen (7) and chromium (2) nearest neighbors (*nn*) and next nearest neighboring (*nnn*) A. The distances are in the range of ~2.5 Å for A-N and ~3 Å for A-Cr. *nnn* A are at > 3.4 Å and it becomes interesting to examine whether bonding can still be traced out then. This is here examined based on the overlap populations ($S_{ij}$) firstly for Ce-N and Cr-N bonding and then for A-Cr and A-A bonding which is pointed out as questionable in literature [5, 6].

Another point which needs addressing is the valence state of cerium, i.e. whether it is trivalent $Ce^{III}$ or tetravalent $Ce^{IV}$. The valence state of Ce is relevant in the magnetic properties of Ce based intermetallic compounds as shown in a review [8]. Ce can be trivalent ($Ce^{III}$), tetravalent ($Ce^{IV}$) or within compounds with intermediate valence cerium. These situations were chemically considered with extended Hückel calculations carried out in the manganese compound $Ce_2MnN_3$ by Niewa et al. [6] who concluded to a tetravalent state. However no magnetic polarization results were provided for Ce/Cr. In the uranium based isostructural compound $U_2CrN_3$, uranium 5f states can behave as 3d transition metals regarding magnetic behavior. Then magnetism can arise from intra-atomic spin polarization (Ce 4f) or interatomic band like (U 5f). Hence, it becomes relevant to examine comparatively the two isostructural compounds for their atom resolved electronic and magnetic properties besides the chemical bonding. We here assess these features based on the density functional theory (DFT) [9, 10].

## 2. Computational methodology

Within the accurate quantum mechanical framework of the DFT we have used two methods in complementary manner.

The Vienna *ab initio* simulation package (VASP) code [11,12] allows geometry optimization, total energy calculations as well as establishing the energy-volume equations of state. The projector augmented wave (PAW) method [12, 13], is used with atomic potentials built within the generalized gradient approximation (GGA) scheme following

Perdew, Burke and Ernzerhof (PBE) [14]. This exchange-correlation XC scheme was preferred to the local density approximation LDA [15] one which is known to be underestimating energy band gaps. The conjugate-gradient algorithm [16] is used in this computational scheme to relax the atoms of the different crystal setups. The tetrahedron method with Blöchl corrections [17] as well as a Methfessel-Paxton [18] scheme was applied for both geometry relaxation and total energy calculations. Brillouin-zone (BZ) integrals were approximated using the special k-point sampling of Monkhorst and Pack [19]. The optimization of the structural parameters was performed until the forces on the atoms were less than 0.02 eV/Å and all stress components less than 0.003 eV/Å$^3$. The calculations are converged at an energy cut-off of 350 eV for the plane-wave basis set with respect to the **k**-point integration with a starting mesh of 6×6×6 up to 12×12×12 for best convergence and relaxation to zero strains.

The charge density issued from the self consistent calculations can be analyzed using the AIM (atoms in molecules theory) approach [20] developed by Bader. Such an analysis can be useful when trends between similar compounds are examined; it does not constitute a tool for evaluating absolute ionizations. Bader's analysis is done using a fast algorithm operating on a charge density grid arising from high precision VASP calculations and generates the total charge associated with each atom.

For a full account of the electron structure, the site projected density of states (PDOS) and the properties of chemical bonding based on overlap matrix ($S_{ij}$) with the COOP criterion [21] within DFT, we used scalar relativistic (needed for post Sn Z=50 elements) full potential augmented spherical wave (ASW) method [22, 23]. The generalized gradient approximation GGA [13] scheme was used to account for the DFT exchange-correlation effects. In the minimal ASW basis set, we chose the outermost shells to represent the valence states for the band calculations and the matrix elements were constructed using partial waves up to $l_{max}$ + 1 = 4 for Ce, U; $l_{max}$ + 1 = 3 for Cr and $l_{max}$ + 1 = 2 for N. Self-consistency was achieved when charge transfers and energy changes between two successive cycles were such as: ΔQ < 10$^{-8}$ and ΔE < 10$^{-6}$ eV, respectively. The BZ integrations were performed using the linear tetrahedron method within the irreducible hexagonal wedge following Blöchl [17].

## 3 Geometry optimization and energy volume equations of states and electron localization

*Geometry optimization.*

Table 1 shows the starting experimental and calculated atomic positions and structure parameters of the two nitrides. A fairly good agreement with experiment can be observed but the internal $z_{N2}$ parameter for $U_2CrN_3$ deviates from experiment while exhibiting a closer value with experimental and calculated magnitudes in the cerium compound. Note however that the experimental data by Benz and Zachariasen [7] assigned the same $z_{N2}$ to the whole series of actinide studied series. For this reason it is more likely to have $z_{N2}$ within range of its value in the cerium compound.

*Energy-volume equation of state.*

Trends of structural properties between the two classes of compounds such as the zero pressure bulk modulus $B_0$, are assessed from the energy (E) - volume (V) values calculated around the optimized *a,b,c* structure parameters (Table 1). $B_0$ expresses the resistance of the material to isotropic compression. From the calculations, the E,V values arrange in E = $f$(V) curves shown in Figs. 2 a,b. The quadratic behavior is indicative of stable minima; i.e. the energy increases on both sides of the E,V minimum. The fits by Birch EOS [24] up to the third order: $E(V) = E_0(V_0) + \frac{9}{8}V_0 B_0 [(V_0/V)^{2/3} - 1]^2 + \frac{9}{16} B_0 (B'-4) V_0 [(V_0/V)^{2/3} - 1]^3$ provides equilibrium parameters: $E_o$, $V_o$, $B_o$ and $B'$ respectively for the energy, the volume, the bulk modulus and its pressure derivative. The obtained values with accurate goodness of fit $\chi^2$ ~$10^{-5}$ / $10^{-4}$ magnitudes are displayed in the insert of Fig. 2 a,b. The equilibrium volume of $Ce_2CrN_3$ of 81 Å$^3$ /FU is closer to experiment than the optimized one (~83 Å$^3$ /FU). Also for $U_2CrN_3$ both optimized and EOS volumes are close and smaller than experiment. The resulting bulk modules are higher than in intermetallics as CeNi with $B_0$ = 77 GPa [25] and within range of oxides and the value for the uranium compound is found close to that of $PdO_2$ [26]. This likely due to the chemical bonding between the metallic elements and nitrogen as discussed in next section. The trend of a smaller $B_0$ for the cerium compound could be partly assigned to the larger volume, however the 50 GPa difference should be also assigned to the nature of the constituents whereby U is harder than Ce.

Based on Bader analysis of charge density within AIM theory presented above, the results of computed charge changes $\Delta Q$ between neutral and ionized elements in the structure are presented in Table 2 for the two compounds.

As expected, charges flow from the metallic elements (Ce, U and Cr) to the non metal (N) at the two nitrogen sites with magnitudes translating little ionic character such as formal $N^{3-}$ Both chemical systems are thus behaving rather covalently. Upon accounting for site multiplicities the total charge change is less than ±5 electrons with slightly larger magnitude in $U_2CrN_3$. From these results one can formally write $[A_2Cr]^{\sim 5+}[N_3]^{\sim 5-}$ thus classifying $A_2CrN_3$ as covalent nitrides, i.e. far from ±9e exchange.

It will be shown here below that detailed electronic and magnetic structure will provide an alternative picture of the valence states.

## 4   Results of electronic density of states and chemical bonding

All electrons full potential scalar relativistic ASW calculations for the electronic band structure and the chemical bonding qualitative analysis were then undertaken.

First, spin degenerate (non-spin polarized NSP) calculations with total spins were performed. This protocol allows to evaluate the tendency of each atomic constituent to magnetic stability/instability through an analysis of the density of states (DOS) at Fermi level ($E_F$): $n(E_F)$ as shown here below. Then spin polarized SP calculations were carried out.

Fig. 3 shows the NSP site projected DOS (PDOS) accounting for the site multiplicities, i.e. 2A, 1Ce, 1 N1 and 2N2. The zero energy along the *x* axis is with respect to the Fermi level $E_F$ which crosses the lower energy part of the Ce(4f) and U(5f) states (resp.) as well as Cr (3d) at finite intensities. The nf PDOS are centered in the empty conduction band CB due to the low filling of both Ce and U 5f subshells already in the atomic state. Nevertheless the crossing occurs at a relatively high PDOS for both nf elements and Cr 3d states. It will be shown that this is connected with instability of the electronic system in such a spin degenerate configuration.

At ~-15 eV the lower energy part of the valence band (VB) shows different behaviors of N1 and N2 s-like PDOS which is also observed from -5 up to $E_F$ for the p block with resulting lower energy N1(PDOS) versus N2(PDOS), pointing out to 'more electronegative' N1. This arises from the different mixing of the two N sublattices with Ce (U) and Cr, ex. Ce states mix preferentially with N2 whereas Cr tendd to mix with N1. Also Cr is more electronegative than Ce and U: $\chi_{Cr}$ = 1.66, $\chi_{Ce}$ = 1.12 and $\chi_U$ = 1.38. However it should be kept in mind the site multiplicities (1 N1 versus 2 N2). Such features can be made more explicit with the analyses of the different interactions thanks to the crystal orbital overlap population COOP criterion based on the overlap integral $S_{ij}$.

In as far as the crystal lattice sites have different multiplicities, i.e. 2A, 1Ce, 1 N1 and 2N2, figs. 4 show the COOP for one of each kind to establish comparison. Positive, negative and zero COOP intensities along y-axis signal bonding, antibonding and non-bonding chemical interactions respectively. In panels a, b showing the metal (C, U, Ce) – nitrogen bonding, most of the VB is found of bonding nature and antibonding states are seen at $E_F$ mainly for Cr-N interaction. Cr-N COOP intensities are larger than Ce-N or U-N, implying that the respective compounds are stabilized by them. This follows from the systematically smaller Cr-N versus Ce(U)-N distances: Cr-N1 ~1.9 Å, Cr-N2 ~ 2.1 Å, Ce(U)-N1 ~2.5 Å, Ce(U)-N2 ~ 2.6 Å.

Panels c, d show the metal-metal bonding. The overall bonding is of much smaller intensity than in panel a, b showing the interaction of the metals with nitrogen which is then the dominant effect in the cohesion of the crystal lattice. Nevertheless the metal-metal bonding is present and shows largest contribution for hetero atomic interaction, i.e. for A-Cr (A = Ce, U), furthermore it is of positive, bonding character throughout the VB especially towards the top of VB; it only starts to be antibonding (negative COOP magnitudes) only above $E_F$ within the CB. This bonding for A-Cr and A-A which was pointed out as questionable in the literature finds an answer here as weak but present.

## 5 Analysis of non magnetic results within Stoner theory and spin polarized SP calculations

The density of states magnitudes at the Fermi level, $n(E_F)$, are provided in Table 2a for the two ternary nitrides. The $n(E_F)$ results can be analyzed within the Stoner mean field theory of band ferromagnetism [27], which merely refers to the onset of magnetic polarization, not to the long range ferromagnetic order. The Stoner theory predicts the system to be unstable in a non-magnetic state if it is characterized by a large $n(E_F)$. The total energy of the spin system results from the exchange and kinetic energies. Referring the total energy to the non-magnetic state (NSP), this is expressed as: E = constant {1- I.n($E_F$)}. In this expression, I (eV) is the Stoner integral and $n(E_F)$ (eV$^{-1}$) is the PDOS value for a given state - here d, f- at $E_F$ in the non-magnetic state. If the unit-less Stoner product I. $n(E_F)$ is larger than 1, energy is lowered and the spin system stabilizes in a (ferro)magnetically ordered configuration. Then the product I.$n(E_F)$ provides a qualitative stability criterion. From literature $I_{Ce}(4f)$ = 0.272 eV [28], $I_U(4f)$ = 0,4488 eV [29] and $I_{Cr}(3d)$ = 0.76 eV [27, 30] –cf. Table 2a. The resulting Stoner products are given at the last lines of Table 2a. They are larger than 1 for Cr in the two compounds as well as for uranium. On the contrary with I.$n_{Ce}(E_F)$= 0.927, Ce does not obey the Stoner criterion for the onset of magnetic moment; however it can be noted the closeness to 1. This rather agrees with a tetravalent state of Ce so that it should not develop a magnetic moment when spin-polarized (SP) calculations are done. Oppositely Cr and U have Stoner products larger than 1 and should develop finite magnetic moments. Also from the magnitude of Cr respective Stoner products, one can expect a larger moment on Cr in the cerium based compound. These expectations need to be further confirmed or infirmed with spin polarized SP calculations.

Scalar relativistic spin-only SP calculations (i.e. without accounting for spin-orbit coupling effects) were started from converged NSP band results. At self convergence of energy and charge transfer the SP configuration was found stabilized versus NSP by -0.53 eV for the cerium compound and -0.62 eV for the uranium compound. Table 2b shows that magnetic moments develop on Ce, Cr and U with magnitudes proportional to the Stoner products. $U_2CrN_3$ has larger cell magnetization than $Ce_2CrN_3$ and it can be noted that this follows the trend of larger energy stabilization of the SP configuration versus NSP through magnetic exchange.

Cerium has a small moment of 0.14 $\mu_B$ while Cr exhibits a large moment of 1.94 $\mu_B$. Ce-Cr overlap of band states, namely the states accounted for in the valence basis set can explain that a residual moment onsets on Ce in spite of a likely tetravalent state; i.e. it is induced by Cr large magnetic polarization. Nevertheless the hypothesis of a contribution from a small amount of Ce$^{III}$ whereby a residual moment on Ce develops cannot be discarded. Nitrogen shows vanishingly small moments at both sites. Partcularly, N1 has a negative induced moment with 4 times larger magnitude than N2 due to the larger Ce-N1 versus Ce-N2 overlap (cf. COOP).

Oppositely to lanthanide element, uranium develops a finite moment close to 1 in $U_2CrN_3$ while Cr has a small moment. Then the Stoner criterion indicator reports correctly the trend towards magnetic polarization while a chemical formal description whereby U and Cr should be trivalent does not stand farther than as descriptive.

Lastly for a search of the magnetic ground state an antiferromagnetic order (SP-AF) was enforced by a double cell in which half the atoms are considered as UP spins and the other half as DONW spin subcell. This was enabled by transforming the structures of the two ternaries from *I* centering to *P* centering leading to double the number of atoms. At self energy convergence there is full compensation of spins and an energy difference ∆E(SP-F – SP-AF) = -0.23 eV / 2 formula units (FU) per cell for the cerium compound and ∆E(SP-F – SP-AF) = -0.11 eV / 2 FU per cell for the uranium compound. Then the predicted long range order of the ground state is ferromagnetic for both ternaries.

For the sake of further illustrating the SP results we show in Fig. 5 the site projected DOS (PDOS) accounting for the site multiplicities in the SP-F ground state.

In $Ce_2CrN_3$ at zero energy along the *x* axis; the Fermi level $E_F$ crosses the lower energy part of the Ce(4*f*) which are mainly present at the bottom of CB. Interesting features appear right at the top of VB where Cr states show similar PDOS to Ce for ↑ states as well as for ↓ states albeit with small intensity. This translates the quantum spin resolved mixing with an energy downshift of ↑ states versus ↓ states, whence the resulting magnetization on Ce. The other states show little energy shift between the majority ↑ and minority ↓ spin populations.

Oppositely within $U_2CrN_3$ the major energy shift can be seen for uranium U(5f) states especially for the 5 time larger $n^{\uparrow}_U(E_F)$ versus $n^{\downarrow}_U(E_F)$ while the Cr PDOS show little energy shift. Tese observation stress further more the magnetization results with small moments on both U and Cr.

The rather broad shape of uranium PDOS versus localized cerium is another illustration of the different behaviors of nf bands in the two systems whereby uranium f behave band-wise like 3d elements oppositely to localized Ce 4f.

REFERENCES

[1] P. Mohn, K. Schwarz, S. Matar, G. Demazeau, Phys. Rev. B, 45 (1992) 4000.

[2] M. Larbi, R. Riane, S. Matar, A. Abdiche, M. Djermouni, M. Ameri, N. Merabet, A. Oualdine A. Z. Naturforsch. B 71 ( 2016) 125.

[3] Mattesini M, Matar S.F., Phys. Rev. B 65 (2002) 075110.

[4] Mattesini M, Matar S.F, Intl. J. Inorg. Mater. 3 (2001) 943.

[5] S. Broll, W. Jeitschko, Z. Naturforsch. Chem. Sci., 50 (1995) 905.

[6] R. Niewa, G. V. Vajenine, F. J. DiSalvo, H. Luo, W. B. Yelon, Z. Naturforsch. 53 b, (1998) 63.

[7] R. Benz and W.H. Zachariasen, J. Nuclear mater. 37 (1970) 109

[8] S.F. Matar, Prog. Solid State Chem., 41, Issue 3, (2013) 55-85 (account).

[9] P. Hohenberg, W. Kohn, Phys. Rev. B 136 (1964) 864.

[10] W. Kohn, L.J. Sham, Phys. Rev. A 140 (1965) 1133.

[11] G. Kresse, J. Furthmüller, Phys. Rev. B 54 (1996) 11169.

[12] G. Kresse, J. Joubert, Phys. Rev. B 59 (1999) 1758.

[13] P. E. Blöchl, Phys. Rev. B 50 (1994) 17953.

[14] J. Perdew, K. Burke, M. Ernzerhof, Phys. Rev. Lett. 77 (1996) 3865.

[15] D. M. Ceperley, B. J. Alder, Phys. Rev. Lett. 45 (1980) 566.

[16] W.H. Press, B.P. Flannery, S.A. Teukolsky, W.T. Vetterling, Numerical Recipes, Cambridge University Press, New York (1986).

[17] P. E. Blöchl, Phys. Rev. B 49 (1994) 16223.

[18] M. Methfessel, A. T. Paxton, Phys. Rev. B 40 (1989) 3616.

[19] H. J. Monkhorst, J. D. Pack, Phys. Rev. B 13 (1976) 5188.

[20] R. Bader, Chem. Rev. 91 (1991) 893.

[21] R. Hoffmann, Angew. Chem. Int. Ed. Engl., 26, (1987) 846.

[22] A. R. Williams, J. Kübler, C. D. Gelatt Jr., Phys. Rev. B, 19,(1979) 6094.

[23] V. Eyert, Int. J. Quantum Chem. 77 (2000) 1007.

[24] F. Birch, J. Geophys. Res. 83 (1978) 1257

[25] S.F. Matar, Solid State Sci. 12 (2010) 59.

[26] S.F. Matar, G. Demazeau, M. H. Möller, R. Pöttgen, Chem. Phys. Lett., 508 (2011) 215

[27] P Mohn. Magnetism in the solid state e an introduction, Springer Series. In: Cardona M, Fulde P, von Klitzing K, Merlin R, Queisser HJ, Störmer H, editors. Solid-state sciences. Heidelberg: Springer; 2003.

[28] S. F. Matar, A. Mavromaras ; JSSC, 49 (2000) 449.

[29] S. F. Matar et V. Eyert, J. Magn. Magn. Mater. 166 (1997) 321.

[30] Janak JF. Phys Rev B 16n(1977) 255

Table 1

a) Experimental [5] and (calculated) crystal data of $Ce_2CrN_3$. Space group *Immm*. Cr and N1 on 2*a*; 0,0,0 and 2b ½,0,0 respectively.

*a* = 3.790 (3.822) Å ; *b* = 3.404 (3.433) Å *c* = 12. 517 (12.623) Å; V(cell)= 161.48 (165.62) Å$^3$

| Atoms (4*i*) | | z |
|---|---|---|
| Ce | 0, 0, z | 0.3537 (0.354) |
| N2 | 0, 0, z | 0.1664 (0.166) |

b) Experimental [6] and (calculated) crystal data of $U_2CrN_3$. Space group *Immm*. Cr and N1 on 2*a*; 0,0,0 and 2b ½,0,0 respectively.

*a* = 3.7397 (3.709) Å; *b* = 3.3082 (3.30)Å; *c* = 12. 3335 (12.181) Å. V(cell)= 152.586 (149.46) Å$^3$

| Atoms (4*i*) | | z |
|---|---|---|
| U | 0, 0, z | 0.356 (0.355) |
| N2 | 0, 0, z | 0.151 (0.169) |

Table 2

a) Stoner integral: **I** (eV) density of states at the Fermi level: $n(E_F)$ in eV$^{-1}$ and unit-less Stoner products I. $n(E_F)$.

|  | $Ce_2CrN_3$ | $U_2CrN_3$ |
|---|---|---|
| I(A) | 0.272 | 0.449 |
| I(Cr) | 0.760 | 0.760 |
| $n_A(E_F)$ | 3.408 | 6.377 |
| $n_{Cr}(E_F)$ | 2.752 | 2.151 |
| I.$n_A(E_F)$ | 0.927 | 2.863 |
| I.$n_{Cr}(E_F)$ | 2.091 | 1.634 |

b) Spin only magnetic moments and cell total magnetization in Bohr magnetons ($\mu_B$)

| $M_{S.O.}/\mu_B$ | $Ce_2CrN_3$ | $U_2CrN_3$ |
|---|---|---|
| M(A) | 0.14 | 0.97 |
| M(Cr) | 1.94 | 0.49 |
| M(N1) | -0.04 | -0.01 |
| M(N2) | -0.01 | -0.02 |
| M(cell) | 2.196 | 2.375 |

**Figures**

Fig. 1: Crystal structure of $A_2CrN_3$ (A= Ce, U)

Fig. 2: Energy volume curves and Birch 3$^{rd}$ order equation of state (EOS) fit values for energy ($E_0$), bulk modulus ($B_0$), volume ($V_0$) and $B_0$-pressure derivative (B') of $Ce_2CrN_3$ and $U_2CrNn$. $\chi^2$ designates the goodness of fit.

Fig. 3: Non spin polarized (NSP; spin degenerate) site projected density of states of a) $Ce_2CrN_3$ and b) $U_2CrN_3$.

Fig. 4: Atom (i)-to-atom (j) bonding magnitudes based on overlap matrix elements ($S_{ij}$) with the COOP criterion. A,c) $Ce_2CrN_3$ and b,d) $U_2CrN_3$.

Fig. 5. Spin polarized (SP; two spin populations' ↑-majority-and ↓ -minority-) site projected density of states of a) $Ce_2CrN_3$ and b) $U_2CrN_3$

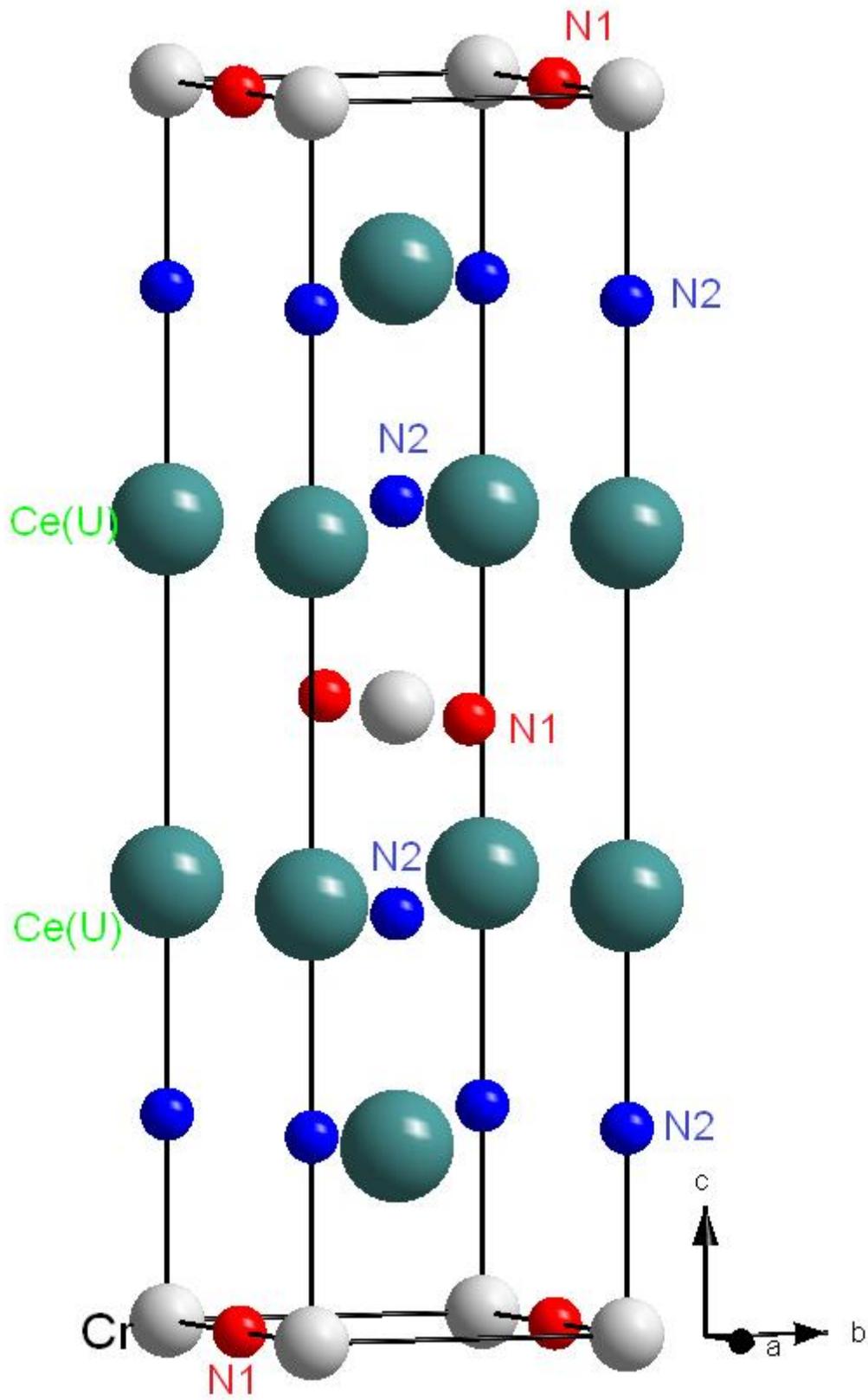

**Fig. 1**

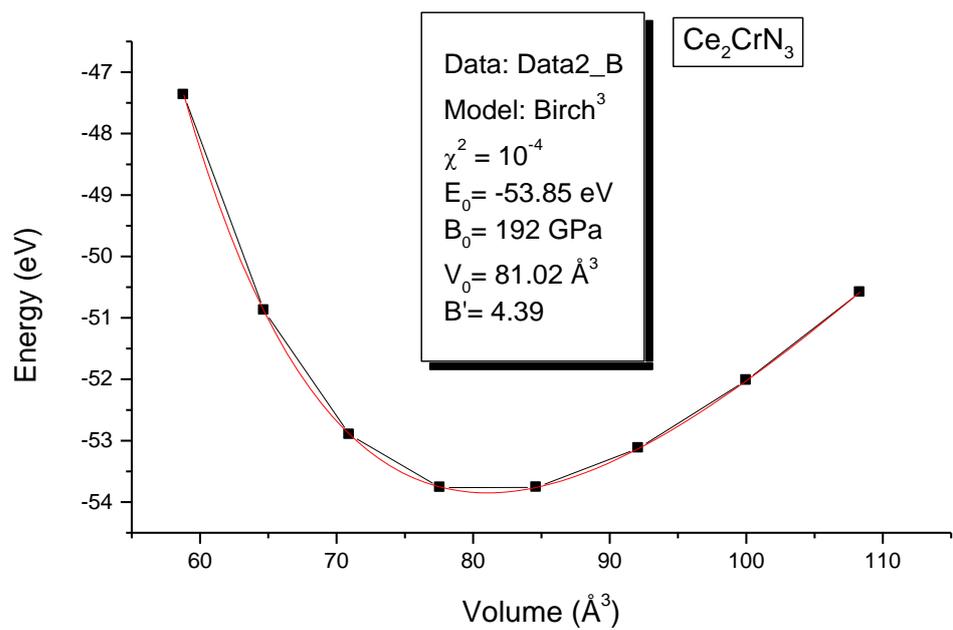

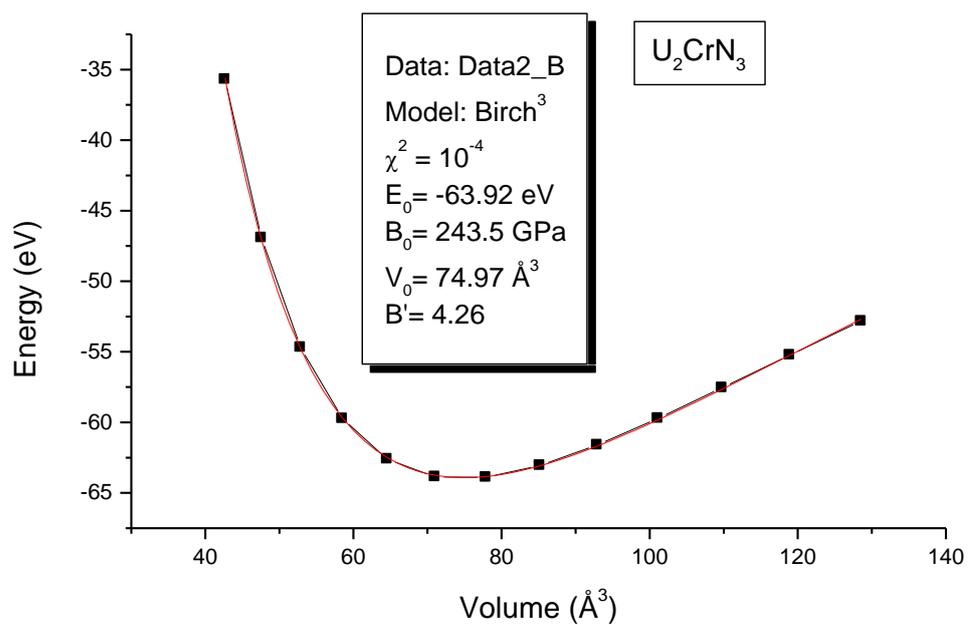

**Fig. 2.**

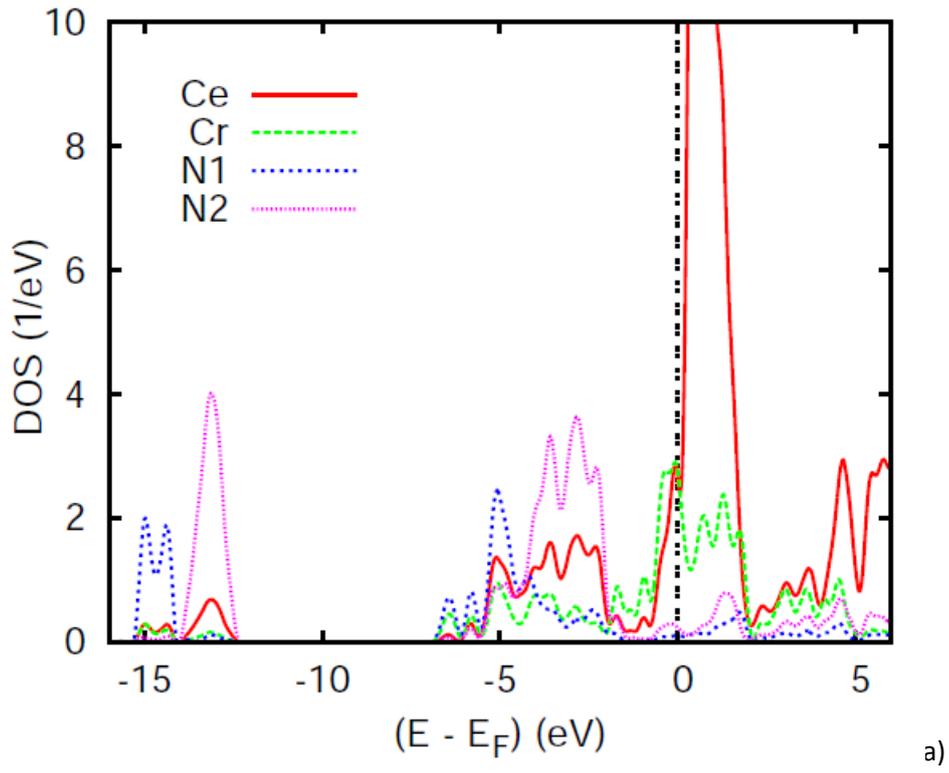

a)

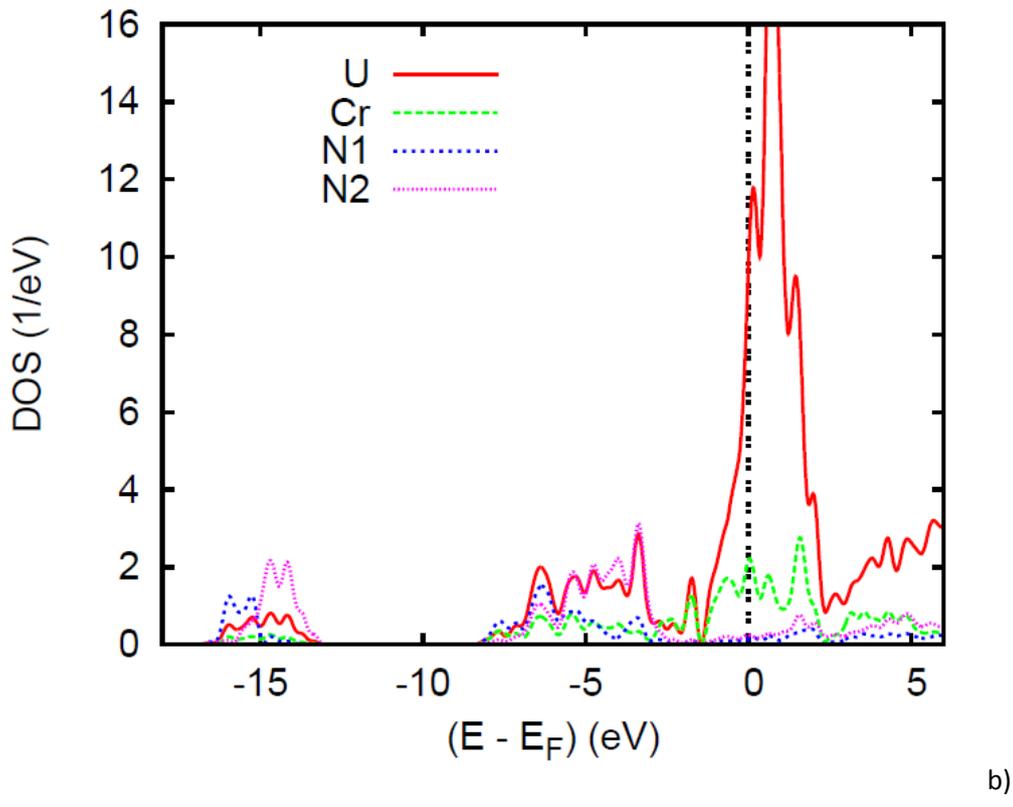

b)

**Fig. 3**

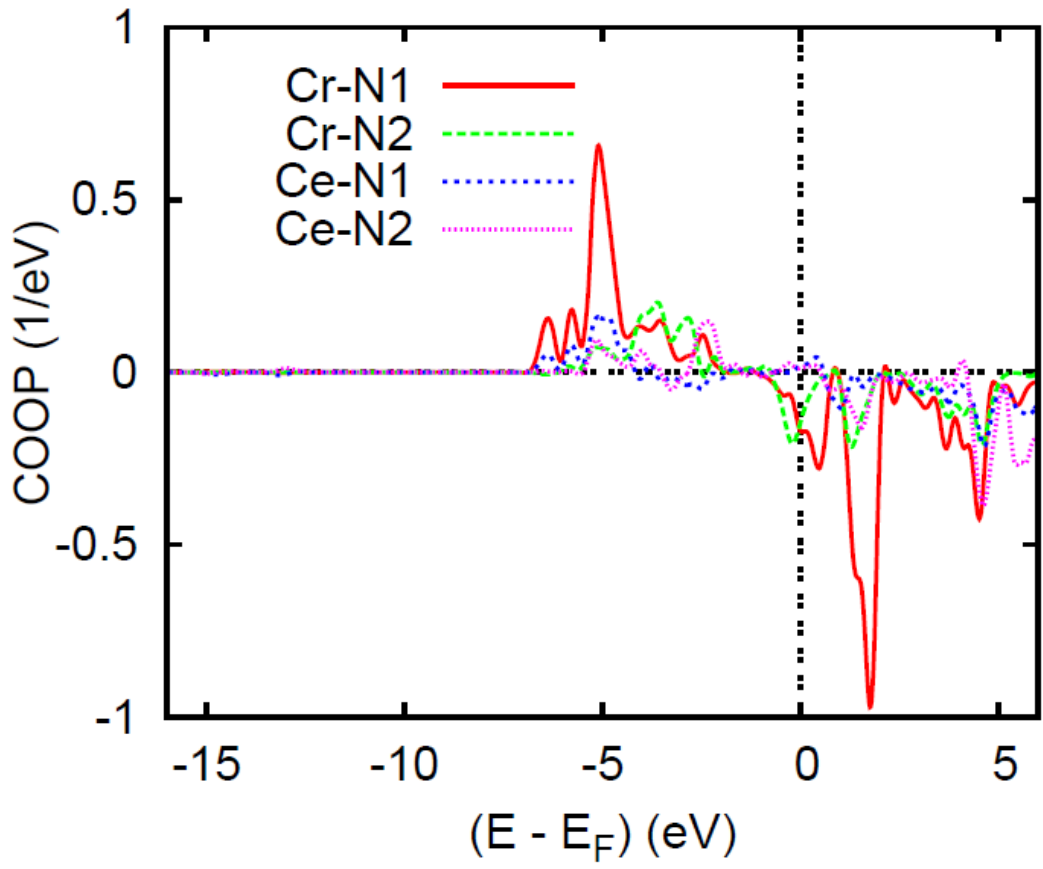

a)

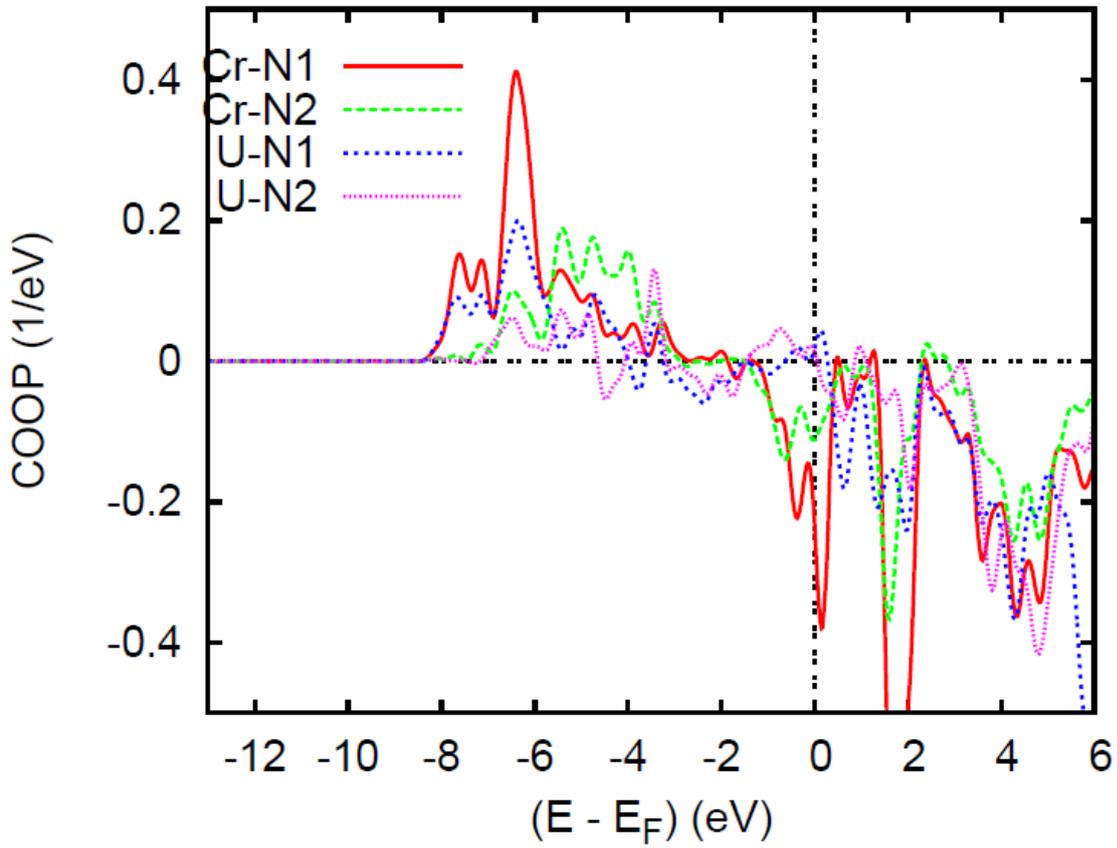

b)

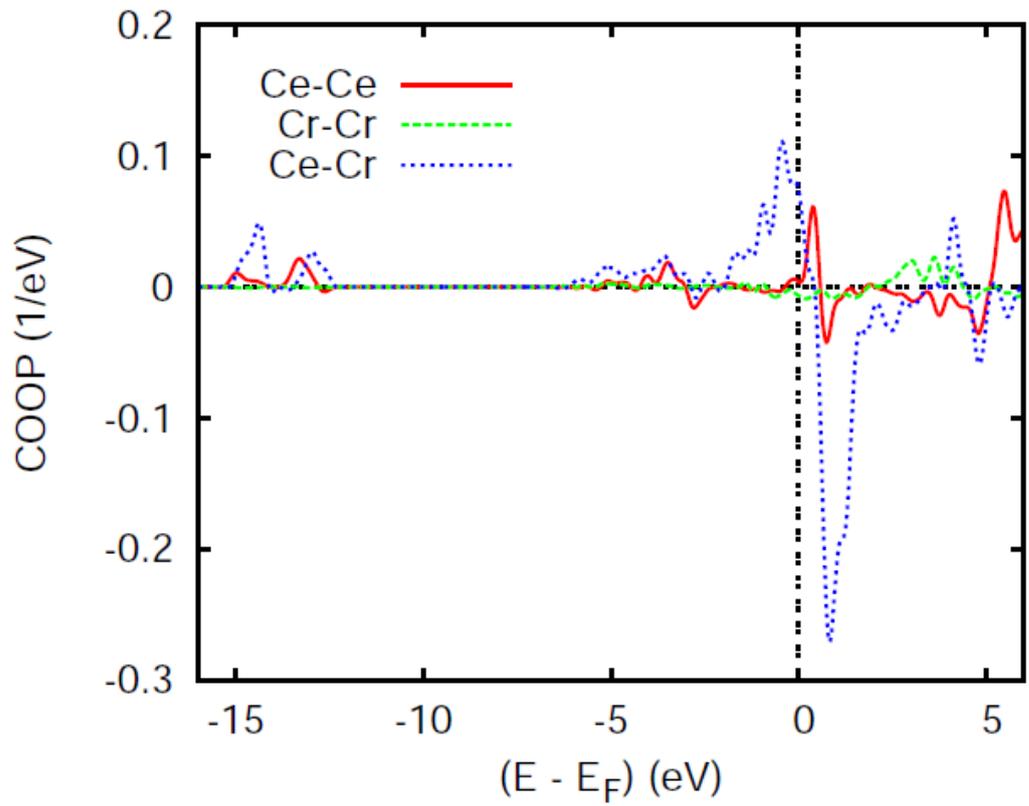

c)

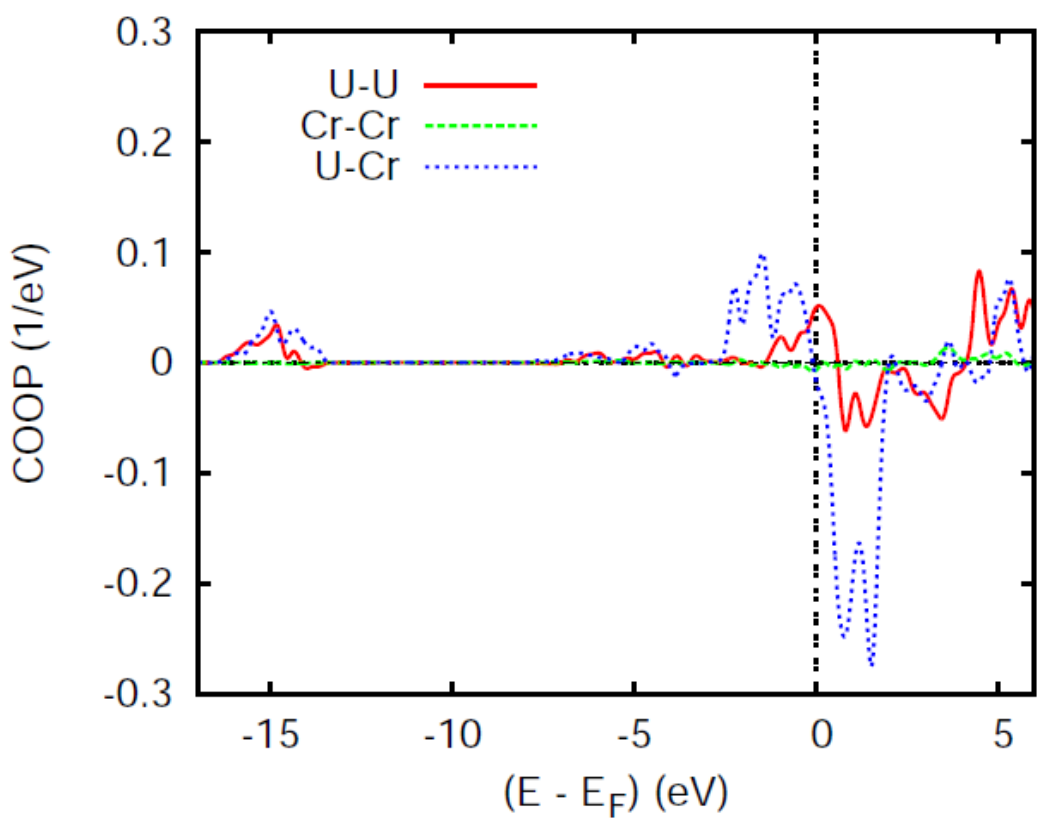

d)

**Fig. 4:**

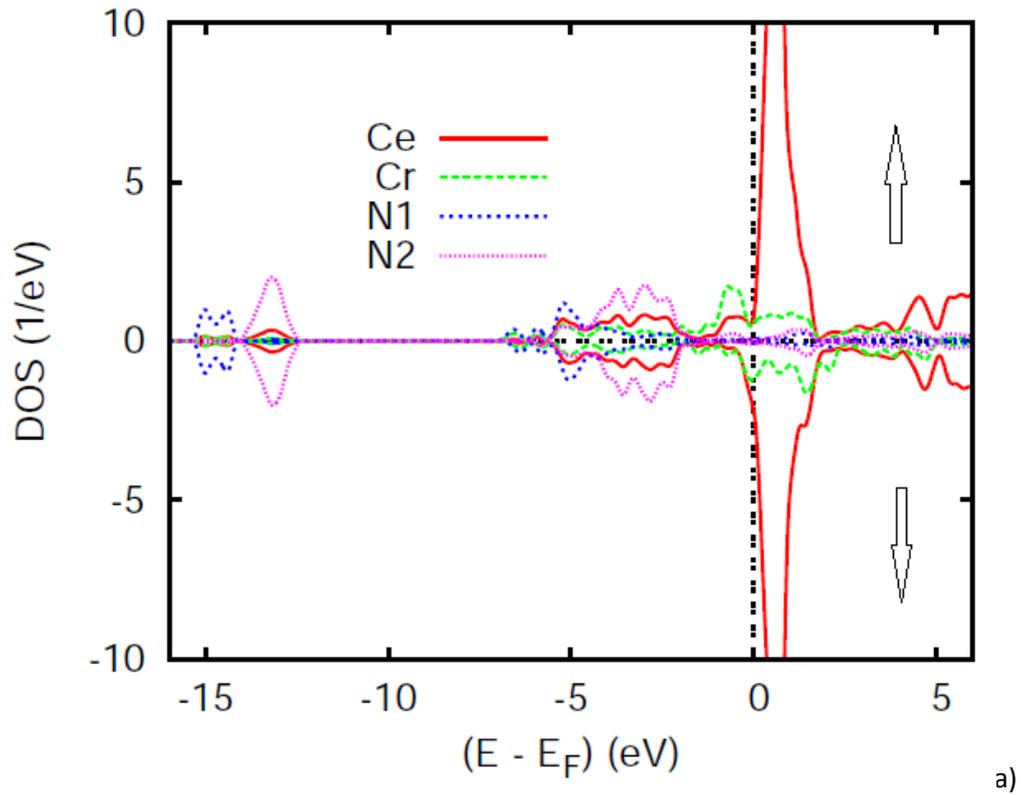

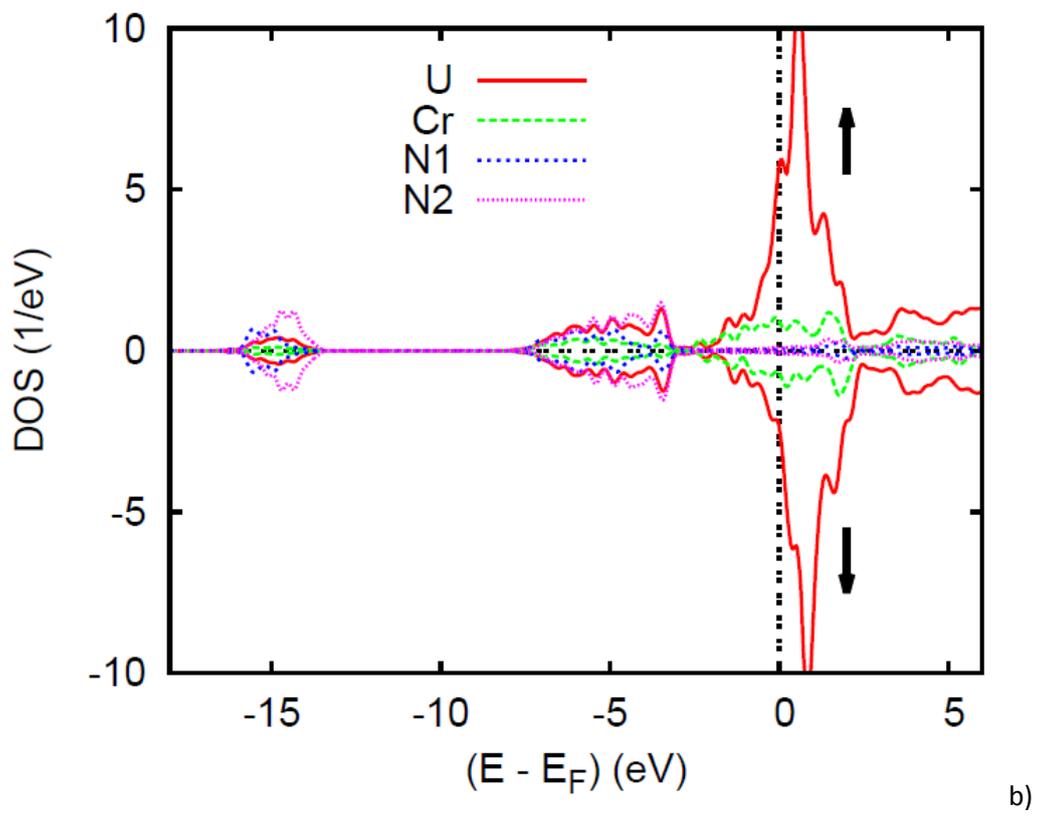

**Fig. 5**